# The Clustering Properties of Faint Galaxies

L. Infante[1,2]
*Grupo de Astrofísica, P. Universidad Católica de Chile,*
*Casilla 104, Santiago 22, Chile*

and

C.J. Pritchet[1]
*Dept. of Physics and Astronomy, University of Victoria,*
*P.O. Box 3055, Victoria, BC, Canada V8W 3P6*

## ABSTRACT

The two-point angular correlation function of galaxies, $w(\theta)$, has been computed from a new survey of faint galaxies covering a 2 deg$^2$ area near the North Galactic Pole. This survey, which is complete to limiting magnitudes $J=24$ and $F=23$, samples angular scales as large as $1.^{\circ}5$.

Faint galaxies are found to be more weakly clustered (by a factor of at least two) compared to galaxies observed locally. Clustering amplitudes are closer to model predictions in the red than in the blue. The weak clustering of faint galaxies cannot be explained by any plausible model of clustering evolution with redshift. However, one possible explanation of the clustering properties of intermediate redshift galaxies is that they resemble those of starburst galaxies and H II region galaxies, which are observed locally to possess weak clustering amplitudes. Our clustering amplitudes are also similar to those of nearby late-type galaxies, which are observed to be more weakly clustered than early-type galaxies

A simple, self-consistent model is presented that predicts the fraction of galaxies in the "excess" population at intermediate redshifts, and correctly matches observed color distributions. The available data on the clustering properties of faint galaxies are consistent with this model if the "excess" population of faint blue galaxies is also the weakly clustered population.

Evidence is presented that the power-law slope of the angular correlation function becomes shallower at fainter magnitudes. A similar effect is seen locally both for dwarf galaxies and for galaxies with late morphological type; this effect is roughly consistent with the model.

*Subject headings:* cosmology – galaxies: clustering – galaxies: evolution

---

# 1. Introduction

## 1.1 Properties of Faint Galaxies

At the Galactic poles, faint galaxies outnumber Galactic stars fainter than about $B = 22$ (Tyson and Jarvis 1979). These galaxies are observed to have mean redshifts of about 0.3 (Colless et al. 1990), corresponding to a lookback time of about 5 Gyr (if $T_o = 15$ Gyr, $q_o = 1/2$). The observed colors of these galaxies are on average slightly *redder* than those of galaxies observed locally; however, after allowing for $k$-corrections, it is clear that a significant fraction (more than 50%) of these galaxies are intrinsically very blue, with rest frame colors as blue as those of Sdm galaxies.

The redshift distributions of faint galaxies (Broadhurst, Ellis, and Shanks 1988; Colless et al. 1990, 1993; Koo and Kron 1992; Lilly 1993) are indistinguishable from the predictions of a no-evolution model. Yet the observed numbers of objects at $J \gtrsim 21$ are significantly in excess of no evolution model predictions (Maddox et al. 1990a). It is also interesting to note that the luminosity function of galaxies at intermediate redshifts ($z \gtrsim 0.2$) contains $2 - 3\times$ as many objects in each magnitude bin around $L^*$ (Eales 1993; Lonsdale and Chokshi 1993; Lilly 1993; Gwyn, Hartwick, and Pritchet 1994). The luminosity density of the Universe is similarly increased at intermediate redshift (Eales 1993; Cowie, Songaila, and Hu 1993).

There appear to be several (related) explanations for the excess counts and luminosity densities at moderate redshifts that are consistent with the observed redshift distributions. Most involve brightening low mass galaxies up to $\sim L^*$. (This evolutionary brightening cannot extend to the most luminous galaxies, because such objects would then be visible at high redshift, and hence would violate the no-evolution form of the redshift distribution.) The excess in the number counts at $J \gtrsim 21$ is probably caused by a population of low mass starburst galaxies that may (Broadhurst et al. 1988) or may not (Cowie, Songaila, and Hu 1991) be visible at the present epoch. (In the former case, the star bursts may be related to the prevalence of intermediate age populations in low mass dwarfs in the Local Group [Hodge 1989]. In the latter case, the present-day counterparts of the excess population at $z \approx 0.3$ might constitute the hidden population of low surface brightness galaxies postulated by Disney [1976; see also Bothun et al. 1990 and McGaugh 1994].)

One possible explanation of the brightening of the low mass galaxies at intermediate redshifts is interactions and mergers, which are known to brighten galaxies locally (e.g. Larson and Tinsley 1978). Plausible merger models have in fact been constructed which are consistent with the excess counts and redshift distributions (e.g. Broadhurst, Ellis, and Glazebrook 1992; Guiderdoni and Rocca-Volmerange 1990; Carlberg and Charlot 1992), and with the numbers of close pairs (Carlberg, Pritchet, and Infante 1994; Colless et al. 1994) (although not with the observed thinness of disks [Tóth and Ostriker 1992] or low density of field ellipticals [Dalcanton 1993]). Note that "vanilla" merger models (with no evolutionary brightening) do not work, because they cannot explain the higher luminosity density of the Universe at intermediate redshifts.

## 1.2 Clustering Properties of Galaxies

A useful statistic for measuring the clustering properties of galaxies is the two point spatial correlation function, $\xi(r)$, which measures the excess probability of finding pairs of galaxies at separation $r$, relative to that expected if galaxies were distributed completely randomly (Peebles 1980, §III). $\xi(r)$ does not of course provide a *complete* description of the galaxy clustering properties of the Universe; it is not sensitive to the detailed topology of structures in the Universe. However, it is one of the simplest statistics computationally, and is related to the power spectrum of density fluctuations (a quantity of theoretical interest) by a simple integral.

The quantity $\xi(r)$ can be computed directly from redshift surveys (e.g. Davis and Peebles 1983; de Lapparent, Kurtz, and Geller 1986; de Lapparent, Geller, and Huchra 1988). Alternately $\xi(r)$ can be inferred from the two point *angular* correlation function $w(\theta)$, which is related to $\xi(r)$ via Limber's (1953, 1954) integral (see also Peebles 1980). Examples of studies of $w(\theta)$ include the Groth and Peebles (1977) analysis of the Lick Survey (Shane and Wirtanen 1967), and, more recently, the Edinburgh/Durham (Collins, Heydon-Dumbleton, and MacGillivray 1989) and APM (Maddox et al. 1990b) surveys.

The principal result from these surveys is that $\xi(r)$ can be represented by a power-law, $\xi(r) = (\frac{r}{r_o})^{-\gamma}$, with correlation length $r_o = 5.4h^{-1}$ Mpc (Davis and Peebles 1983) and $\gamma$ often quoted as 1.8 (cf. Groth and Peebles 1977, Davis and Peebles 1983, who find $\gamma = 1.77$). Other vales of $r_0$ appear in the literature (cf. $4.7h^{-1}$ Mpc – Groth and Peebles 1977; 7.5 $h^{-1}$ Mpc – de Lapparent, Geller, and Huchra 1988; 6.2 $h^{-1}$ Mpc – Vogeley et al. 1992), with the larger values coming from redshift surveys which can be affected by redshift-space distortions (e.g. Kaiser 1987, Fry and Gaztañaga 1994). (In this paper we adopt the value $r_o = 5.4h^{-1}$ Mpc for comparison with the high redshift sample.) There is evidence for a break at large $r$, first noted by Groth and Peebles (1977). The reality of this break was challenged by several groups (e.g. Geller, de Lapparent, and Kurtz 1984); recent survey work does in fact indicate the presence of a break, but the scale of the break remains controversial (cf. Maddox et al. 1990b, Collins et al. 1989).

The original motivation for studying the clustering properties of *faint* galaxies was to search for evidence of evolution in the correlation function, and to provide a constraint on galaxy (i.e. luminosity) evolution (Koo and Szalay 1984). More recently it has become clear that the clustering properties of faint galaxies provide important information on the nature of the "faint blue galaxy" population at intermediate redshift. Studies of the angular correlation function of faint galaxies include those of Ellis (1980), Koo and Szalay (1984), Stevenson et al. (1985), Pritchet and Infante (1986), and Jones et al. (1987). More recent work, which will be discussed below, includes that of Efstathiou et al. (1991), Pritchet and Infante (1992), Neuschaefer et al. (1991), Bernstein et al. (1994), Couch, Jurcevic, and Boyle (1993), Roche et al. (1993), and Brainerd, Smail, and Mould (1994).

### 1.3 This Paper

In this paper we study the angular correlation function of faint galaxies using a catalog derived from $\sim 2$ deg$^2$ of photographic material by Infante and Pritchet (1992, hereafter Paper I). The principal result is that faint galaxies are intrinsically more weakly clustered than galaxies at the present epoch. This result, which largely negates the possibility of using faint galaxy clustering as a tracer of galaxy evolution, was presented in preliminary form in an earlier paper (Pritchet and Infante 1992); the present paper provides more details on the data analysis, and also on the interpretation of the low correlation amplitudes of faint galaxies.

The properties of the catalog (derived in Paper I) are summarized in §2. Algorithms that were used for determining the angular correlation function are briefly discussed in §3 (see Infante 1994a for further details). §4 presents the angular correlation function measurements, and §5 presents the interpretation of these observations (see also Pritchet and Infante 1992).

## 2. The Catalog

The following is a brief summary of the steps that were taken in obtaining and calibrating the data. The reader is referred to Paper I for more details.

### 2.1 Observations

In 1987 April the Canada-France-Hawaii 3.6m telescope was used to obtain 9 plates (5 in $J$ and 4 in $F$) of 5 adjoining fields near the North Galactic Pole. The fields were positioned in the overlapping chequerboard distribution shown in Figure 1; from extensive simulations in Paper I, this layout was found to provide optimal angular coverage and lowest noise in measuring $w(\theta)$. The emulsion/filter combinations were IIIaJ+GG385 = Kron (1980) $J$, and IIIaF+GG495 = Kron $F$. The seeing for all observations was $\lesssim 1''$ FWHM. The average scale of the CFHT prime focus plane is 13.64 $''$ mm$^{-1}$, and the unvignetted field per plate was 0.84 deg$^2$.

### 2.2 Digitization and Parametrization

The plates were scanned with the Automatic Plate Machine [APM] in Cambridge, England; photometric and structural parameters were derived for all stars and galaxies using on-line software. The central regions of each plate were rescanned in raster mode; the density rasters from these scans were used in the photometric corrections described in §2.4 below.

### 2.3 Classification

Star/galaxy discrimination, and further galaxy subclassification, were performed using standard APM software. Parameters derived from the APM moments were plotted against each other; the principal diagnostics of image structure were found to be plots of area vs. magnitude, mean size vs. magnitude and size vs. peak surface brightness. The probability of an object being a star or a galaxy was assigned by manually defining the class boundary. Classification was found to be very reliable for $J \lesssim 23$; for fainter objects classification is not critical because galaxies outnumber stars by a large factor (typically 10 to 1) at the galactic poles. (The stars are discussed further in Infante [1994b].)

## 2.4 Total photometry

In order to transform instrumental APM *isophotal* magnitudes to linear *total* magnitides (*e.g.* Kron 1980, Irwin and Hall 1983), the central raster scans for each plate were first converted to intensity using a density-to-intensity calibration curve constructed from stars, and then analyzed with a *total* magnitude algorithm (Irwin 1988). A conversion from isophotal to total magnitudes was then determined separately for each class of image and for each field. The isophotal APM magnitude was then corrected to a total magnitude for every star and galaxy in the APM catalog.

## 2.5 Zero Points

Sensitivity and, to a lesser extent, scale variations across the plates result in radially varying zero points. $B$ and $R$ CCD standard sequences were therefore established for five locations on each plate in order to monitor these variations. The change in zero point from the center to the edge of the plate amounts, in the worst case, to 0.2 mag; this effect has been corrected. The error in the determination of the zero points at the center of the plates is typically $\lesssim 0.03$ mag. Anticipating a result from §3, it is predicted that systematic errors of $\pm 0.03$ mag in zero points result in errors of about $\pm 0.0008$ in $w(\theta)$.

## 2.6 Coordinate Transformations

In order to transform machine coordinates to celestial coordinates, secondary astrometric standards were obtained from a red POSS plate (O1393). This plate was scanned with the APM, and nearly 2500 stars were identified. $x, y$ coordinates of these stars were transformed to $\alpha, \delta$ using SAO standards. Two fifth order polynomials in $x$ and $y$, each with 21 terms, were then fitted to CFHT observations of the POSS secondary astrometric standards to define a transformation of CFHT $x, y$ to the SAO $\alpha, \delta$ system. The final coordinates were found to have an internal error of smaller than 0.5 arcsec.

## 2.7 Editing the Catalog

A variety of bad images and contaminated areas were removed from the catalog. These included: *(i)* multiple hits on bright stars and galaxies, *(ii)* images contaminated by plate defects, *(iii)* images fainter than $J = 24$ and $F = 23$, *(iv)* entire areas showing obvious defects such as diffraction spikes, double images, bright stars, etc., and *(v)* vignetted areas (see Figure 1). Points inside the contours and outside the masked areas are included for final analysis of the catalog.

A total of 39543 galaxies were detected in the $J$ catalog, and 38912 galaxies in the $F$ catalog. The two catalogs were also merged to produce a list of 21474 galaxies that were present in *both* $J$ and $F$ catalogs; this merged catalog is referred to as the $M$ catalog. The catalogs cover an effective area of about 2 deg$^2$ (after masking of bad areas), and are complete to limiting magnitudes of $J \simeq 24$ and $F \simeq 23$. The excellent seeing for our CFHT plates ($\lesssim 1''$ FWHM), and the careful monitoring of zero points across the plates, make this catalog especially well suited to studies of large scale structure.

## 3. Estimation Of The Angular Correlation Function

The two-point angular correlation function of galaxies, $w_{gg}(\theta)$, measures the mean *excess* surface density of pairs of galaxies centered on some reference galaxy, compared to what would have been measured if galaxies were distributed randomly on the sky. Techniques for estimating $w_{gg}(\theta)$ are discussed in detail by, for example, Sharp (1979), Hewett (1982), Pritchet and Infante (1986), and Infante (1994a, Paper II). Our techniques for calculating $w(\theta)$ differ in some details from those used by other groups, and are discussed in detail in Paper II, to which the reader is referred. Paper II also describes simulations of clustered data that were used to test the algorithms that are summarized below.

It is well known that the angular auto-correlation function can be estimated by (Peebles 1980)

$$1 + w_{gg}(\theta) = \frac{2N_{gg}(\theta)\Omega}{N(N-1)\langle\delta\Omega\rangle}, \quad (1)$$

where N is the number of objects in the survey, $N_{gg}$ is the number of galaxy pairs between separations $\theta$ to $\theta + \delta\theta$, $\Omega$ is the solid angle subtended by the area being surveyed, and $\langle\delta\Omega\rangle$ is the mean value of the solid angle subtended by annuli (with inner and outer radii $\theta$ and $\theta + \delta\theta$) that are centered on the catalog points. (In practice $\langle\delta\Omega\rangle$ is affected by the catalog boundaries, and by areas of incompleteness within the catalog boundaries.) Similarly, the estimate of the angular *cross* correlation function of a galaxy sample $g$ with respect to a random sample $r$ is given by

$$1 + w_{gr}(\theta) = \frac{N_{gr}(\theta)\Omega}{N_g N_r \langle \delta\Omega \rangle} , \qquad (2)$$

where $N_{gr}$ is the number of pairs between separations $\theta$ to $\theta + \delta\theta$ when the points in the *galaxy* catalog are taken as centers.

The principal difficulty in estimating $w(\theta)$ from a catalog with irregular boundaries is in the calculation of $\langle \delta\Omega \rangle /\Omega$. This quantity represents the mean fractional solid angle subtended by the separations $\theta$ and $\theta + \delta\theta$. In the case where no boundaries are present, the mean solid angle of the ring is simply given by $2\pi[\cos(\theta) - \cos(\theta + \Delta\theta)]$. However, the presence of edges complicates the computation of $\langle \delta\theta \rangle$ (*e.g.* Peebles 1980, Sharp 1979, Hewett 1982); in this paper we use Monte Carlo techniques to evaluate edge effects, as follows. $N_r$ random points are placed within a solid angle $\Omega$; these points may be thought of as a "tracer" population. Taking galaxies as centers, the number of random points are computed at separations between $\theta$ and $\theta + \delta\theta$ from the galaxies; this result will be called $N_{gr}$. Since $w_{gr}(\theta) = 0$ for non-correlated samples, it follows that

$$\frac{\langle \delta\Omega \rangle}{\Omega} = \frac{N_{gr}}{N_g N_r} . \qquad (3)$$

Substituting this equation in equation 1, an estimate of $w(\theta)$ may then be written

$$1 + w(\theta) = \frac{2 N_{gg} N_r}{N_{gr}(N_g - 1)}. \qquad (4)$$

In fact, the full formula that is used to estimate $w(\theta)$ in this paper is

$$1 + w(\theta) = \frac{N_{gg} N_r}{B N_{gr}(N_g - 1)} - w_{rg}(\theta), \qquad (5)$$

where $B$ and $w_{rg}(\theta)$ are corrections that are discussed in detail below.

In previous work the angular correlation function for faint samples of galaxies had been estimated using $N_{rr}$ instead of $N_{gr}$ in equation 5 (*e.g.* Ellis 1980, Koo and Szalay 1984, Pritchet and Infante 1986, Jones *et al.* 1987), with a further scaling by a factor $(N_r - 1)/N_g$ to correct for the use of this term. Simulations described in Paper II show that the use of $N_{rr}$ instead of $N_{gr}$ above can result in spurious "humps" in $w(\theta)$ at large separations, and is not to be recommended. These errors are most likely produced by the largest–scale (lowest spatial frequency) variations in galaxy density relative to the field boundaries.

As discussed in Paper II, three corrections must be made to the estimate of $w(\theta)$. Two of these corrections (the integral constraint, $B$, and the random-galaxy cross correlation, $w_{rg}(\theta)$) appear in equation 5; the third (contamination by stars) does not have a strong effect at faint magnitudes. We discuss each of these corrections in detail below.

*1. Random-galaxy Cross Correlation.* This quantity can be used to correct for any spurious correlation introduced by the large-scale distribution of galaxies relative to the field boundaries. Its use is based upon the suggestion of Hewett (1982); $w_{rg}(\theta)$ is calculated from

$$1 + w_{rg}(\theta) = \frac{N_{rg}(\theta)(N_r - 1)}{2 N_{rr}(\theta) N_g} , \qquad (6)$$

where $N_{rg}$ is the number of pairs obtained using the points in the *random* catalog as centers.

*2. Integral constraint.* For a bounded region, the mean galaxy density from equation 1 is effectively $N_g/\Omega$, which is an overestimate because of positive correlations at small separations. To correct for this, a factor $B$ is included in equation 5. This factor, which was estimated by trial and error in some previous work, or by assuming a value for the slope, has a very strong effect on the value of $w(\theta)$ at the largest separations, and, if computed incorrectly, can cause spurious cutoffs to appear at large $\theta$.

A new method for computing $B$ has been developed in Paper II. The correct value of $B$ *must* be consistent with the final computed $w(\theta)$. To enforce this consistency $B$ can be estimated iteratively from the following expression:

$$B = \frac{\Sigma_i N_{rr}(\theta_i)}{\Sigma_i (1 + w(\theta_i)) N_{rr}(\theta_i)} , \qquad (7)$$

where $\Sigma_i$ is a summation over all annular zones ($\theta_i$ bins), and $N_{rr}$ is obtained using a random sample covering the field of interest. In practice, one starts with the assumption $B = 1$, and computes $w(\theta_i)$ using equation 5. $B$ is then estimated from equation 7; new values for $w(\theta_i)$ are computed, an improved value of $B$ is estimated, and a final set of $w(\theta_i)$ values are obtained. Four (or fewer) iterations usually suffice to converge on values of $B$ and $w(\theta_i)$ that are self-consistent. A detailed discussion of this technique appears in Paper II.

Formally, this technique for estimating $B$ assumes that the survey is large enough in angular extent to reach scales on which there is no power in the angular correlation function. Our survey reaches angular scales of order 1°, as can be seen from the pair distribution in Fig. 2. As we shall show, $w(\theta)$ is very small or negligible at faint magnitudes for $\theta \gtrsim 1°$. An extrapolation of our correlation functions beyond 1° predicts that our $B$ factors should be unaffected by missed correlation power at the >99.8% level – i.e. our correlation functions will be underestimated by < 0.002 due to the effects of large scale power on our integral constraint algorithm.

*3. Contamination by Stars.* It is straightforward to show that contamination of a galaxy catalog by *uncorrelated* objects such as stars (due to, for example, faulty star/galaxy discrimination) reduces the amplitude of the correlation function. If a fraction $f$ of the catalog objects is stars, then the amplitude of the correlation function is reduced by a factor $(1-f)^2$ *at all separations*, but the *shape* of the correlation function is unaffected.

Contamination by stars produces negligible errors in the present study. At faint magnitudes, galaxies outnumber stars by a large factor at the Galactic poles; for example, at $J = 23$ (24) only 15% (10%) of all objects are stars (Kron 1980). It follows that the effect on the amplitude of $w(\theta)$ is quite small. At brighter magnitudes stars represent a greater fraction of all objects in the sky, but star/galaxy discrimination is considerably more effective because brighter galaxies are larger and more easily resolved.

Non-systematic errors in the estimate of the angular correlation function are dominated by Poisson noise, by small variations in the magnitude zero points across the fields, and by variations in the limiting magnitude of the sample (due to, for example, variations in image quality across the plates). The Poisson error in $w(\theta)$ at each bin between $\theta$ and $\theta + \delta\theta$ is expected to be (Peebles 1980),

$$\delta w(\theta) = \sqrt{\frac{1+w(\theta)}{N_{gg}}}, \qquad (8)$$

which, for small correlation amplitudes (large angles), reduces to standard $\sqrt{N}$ statistics. Throughout this work the $\pm 1\sigma$ error bars in the log $w$ vs. log $\theta$ diagrams represent the above uncertainty.

For catalogs with large numbers of objects, and at separations with large numbers of pairs of objects, the above statistical uncertainty is unimportant compared to errors due to variations of photometric zero points across the sampled area (Geller, de Lapparent, and Kurtz 1984; deLapparent, Kurtz, and Geller 1986). The effect of a magnitude error $\Delta m$ (in either the zero point or limiting magnitude of the sample) on $w(\theta)$ is given by

$$\delta w(\theta) \approx [0.4 \log_e(10) \Delta m]^2. \qquad (9)$$

Here it is assumed that galaxy counts vary with magnitude as log $N \propto 0.4\ m$, where N is the surface density of galaxies per unit solid angle, and $m$ is magnitude. (The logarithmic slope d log(N)/dm $\approx 0.4$ is a reasonable compromise for galaxy counts in the $J$ and $F$ bands – e.g. Paper I.) The "noise" produced in $w(\theta)$ by random variations in zero point is independent of separation $\theta$.

The variation in limiting magnitude is difficult to quantify in this catalog. Because of the excellent seeing conditions under which these plates were taken, the data appears to be > 50% complete at $J = 24.5$ and $F = 23.5$. We have taken a "conservative" cut of the data at $J = 24$ and $F = 23$ and hence should have minimized the effects of variations in limiting magnitude. Plate-to-plate differences in $m_{lim}$ have little effect on this work, both because of the above, and also because all conclusions that were drawn from the ensemble of plates were also verified from individual plates. The reader is referred to Paper I for further details.

The pair distribution for our survey is shown in Fig. 2. For virtually all separations, the numbers of pairs in the catalog are very large, and so the effects of zero point variations (eqn. 9) dominate over those of Poisson noise (eqn. 8), at least for magnitude errors $\gtrsim 0.01$ mag. However, there is also noise introduced into our estimate of the correlation function by the irregular boundaries of the fields. From the simulations in Paper II we estimate this latter source of noise to be approximately 5–10% in $w(\theta)$ at $\theta \simeq 0°.1$ (rising at smaller and larger separations). Clearly this effect dominates over other sources of noise, except at very faint magnitudes and at large separations. Anticipating results from §§4.1 and 4.2, these errors can be compared with the *measured* rms errors (from plate-to-plate differences) in $w(\theta)$ of $\pm 17\%$ in $J$ and $\pm 12\%$ in $F$ for the full field, roughly double what the simulations predicted. This could be a manifestation of plate-to-plate variation in the magnitude zeropoint or limiting magnitude at the 0.05–0.1 mag level.

## 4. Results

Using the procedures outlined in §3, $w(\theta)$ was estimated as a function of *(i)* limiting magnitude in the $J$ and $F$ catalogs, *(ii)* limiting magnitude for individual fields in the $J$ and $F$ catalogs, and *(iii)* color in the combined $J$ and $F$ catalogs (the merged or $M$ catalog). (We do not discuss the correlations obtained as a function of $(J+F)/2$ from the $M$ catalog in much detail, because they are not completely independent of the results for $J$ and $F$.)

The $w(\theta)$ data were in all cases fitted to a power-law in $\theta$ with an outer cutoff (*i.e.* a linear segment with an outer cutoff in the log $w$ vs. log $\theta$ plane) using weighted least squares. The specific equation that was fitted to the data was

$$w(\theta) = A_w \theta^{-\delta}[1 - \left(\frac{\theta}{\theta_c}\right)^{-\beta}]. \qquad (10)$$

This provides a standard power-law form for $w(\theta)$ at small separations, $w(\theta) = A_w \theta^{-\delta}$, with a cutoff $\theta_c$, defined as the point where $w(\theta)$ reaches zero. $\beta$ is a large number, *e.g.* $\beta = 15$, to make the cutoff sharp. Note that in many previous papers the normalization of $A_w$ was taken at $\theta = 1°$; in this paper we use $A_w = w(1')$.

The cutoff angle is in practice very poorly determined because of the small amplitude, and concomitant huge errors, in $w(\theta)$ at large $\theta$, where the cutoff term starts to dominate; however, the inclusion of this term in equation 10 is necessary to fit $w(\theta)$, as will be shown below. A more robust parameter appears to be the "break angle", $\theta_b$, at which $w(\theta)$ departs from the power law that was fitted at small angles; this quantity is determined from visual inspection. In practice it appears possible to determine $\log \theta_b$ consistently to within $\pm 0.1$ (*i.e.* to within $\pm 25\%$ in $\theta_b$).

It should be borne in mind that the fitting parameters discussed above were chosen for the sole purpose of providing a quantitative description of the shape and amplitude of $w(\theta)$. They were also found to be helpful in studying variations in $w(\theta)$ among different subsets of the data. Nevertheless, the manner in which the cutoff is introduced in eqn. 10 is completely *ad hoc*. A more physical fitting function might be obtained, for example, from the family of functions that Vogeley *et al.* (1992) use to fit power spectra. However, this would make little difference to any of the conclusions drawn in this paper because the cutoff is so difficult to measure reliably. Furthermore, the break that we see in our correlation functions occurs at a physical scale $\sim 5h^{-1}$ Mpc, which can be compared with the break in the correlation function of lower $z$ galaxies in the APM survey ($\sim 20$ Mpc). Our interpretation of this is that the observed breaks in our work may be due to a (small) overestimate of the integral constraint factor $B$ (*e.g.* §3), although we cannot exclude the possibility that they are real. We include the break and cutoff angles in Tables 1 and 3, but do not discuss these results further.

Finally, we determined the amplitudes $A_w^{\delta=0.8}$, $A_w^{\delta=0.66}$, and $A_w^{\delta=0.6}$ of lines that were forced to fit the logarithmic data with $\delta = 0.80, 0.66,$ and $0.60$. (The value $\delta = 0.8$ is of course the canonical slope of the correlation function assumed by many groups [*cf.* Groth and Peebles 1977]. $\delta = 0.66$ is the value determined from the APM survey [Maddox *et al.* 1990b], and $\delta = 0.60$ corresponds roughly to the flattest slopes found in this paper.) These fits were performed for $0°.001 \leq \theta \leq 0°.1$.

We now turn to a detailed discussion of the observed $w(\theta)$ relations for our NGP fields.

### 4.1 J Catalog Correlations

The correlation function for the $J$ catalog is shown in Figures 3 to 5; fitted parameters may be found in Tables 1 (fit of both $A_w$ and $\delta$ in eqn. 10) and 2 (fit assuming fixed $\delta$). The dashed line in the figures represents a power law with a slope $\delta = 0.8$ and an amplitude $A_w = 0.080$ at a scale of 1 arcmin (equivalent to $A_w = 0.003$ at a scale of 1 degree). The error bars represent the $\pm 1\sigma$ uncertainty from equation 8. No attempt has been made to include the other sources of error that were discussed in §3. (The reader is referred to Paper II for a full discussion of statistical and systematic uncertainties in the estimation of $w(\theta)$.)

Figures 3 and 4 demonstrate that roughly similar effects are seen in data subsets defined by limiting magnitude and by magnitude range. This is not a surprising result, since, given the steep exponential slope of galaxy number counts (d log N /dm $\approx 0.4$), most ($\sim 60\%$) of galaxies in a magnitude-limited sample are in the faintest 1 mag bin; the median magnitude of a magnitude-limited sample of galaxies is about $m_{lim} - 0.75$. This implies that $w(\theta)$ determined for all galaxies brighter than some magnitude limit will not differ substantially from $w(\theta)$ defined for the faintest 1 mag subset of the same catalog.

Figure 5 presents a comparison of $w(\theta)$ estimates for the 5 individual fields ($J < 24$). It can be seen that the individual fields are in reasonable agreement, with the exception of data at small separations ($\theta \lesssim 10$ arcsec). The reason for this discrepancy is unclear, but from careful inspection of the data we have found independent evidence that some objects at small separations may be missing from the APM data. (A reanalysis of $w(\theta)$ at small separations will appear in McCracken, Pritchet, and Infante [1995].) For $\theta > 10''$, the field-to-field rms scatter in log $w(\theta)$ among the fields averages $\pm 0.15$, or $\pm 0.07$ in the mean ($\pm 17\%$ in $w(\theta)$). We regard this as a reasonable estimate of the external error in our $w(\theta)$ estimates averaged over all fields (*cf.* §4). Most importantly, the figure shows that, with the possible exception of Field #3, $w(\theta)$ has the same general shape (slope, amplitude, and "break" from a power law) for the individual fields. A similar result is found for the $J < 23$ and $J < 22$ subsamples.

Table 1 shows that the mean slope of the angular correlation function flattens systematically from the brightest ($21 < J < 22$) to the faintest ($23 < J < 24$) magnitude bins. The decrease in $\delta$ with increasing magnitude appears to be significant at a $> 95\%$ level (*i.e.* the hypothesis that $\delta$ is constant can be formally rejected at this confidence level). The effect is also clearly visible in Figures 3 and 4. Furthermore, the flattening of $\delta$ is seen in the data for individual fields, for which mean power law slopes of $0.63 \pm 0.07$ ($J < 24$), $0.78 \pm 0.10$ ($J < 23$), and $0.77 \pm 0.05$ ($J < 22$) are obtained. (Here the $1\sigma$ errors are from the fit, but these are also consistent with the scatter of slopes among the 5 separate fields.) This result is consistent with *some* previous work, and is discussed further in §5. It is hard to imagine ways in which systematic errors could cause such a flattening in the power-law slope, and the effect is not visible in any of the simulations that were performed in Paper II.

The mean amplitude $A_w^{\delta=0.8}(1')$ of $w(\theta)$ is obtained by fitting a power-law with $\delta = 0.8$ to the data at $\theta \leq 0°.1$ (Table 2). Using the full $J$ catalog, this quantity is found to decrease by a factor of order 3 between $J = 22$ and $= 24$. A similar effect is also seen in the 5 individual fields which comprise the $J$ catalog: mean $A_w^{\delta=0.8}(1')$ values of $0.032 \pm 0.005$ ($J < 24$), $0.043 \pm 0.005$ ($J < 23$), and $0.124 \pm 0.037$ ($J < 22$) are found for these 5 fields when fit individually ($1\sigma$ errors from the fit). Given the errors, these values are consistent with the those obtained from the analysis of the full area (Table 2).

### 4.2 F Catalog Correlations

The present $F$ catalog is complete to a limiting magnitude $F = 23$. The surface density of galaxies at this limit is approximately the same as for the $J$ catalog, which possesses a limiting magnitude $J = 24$; however, as pointed out in Paper I, only about half of the galaxies brighter than $F = 23$ possess $J$ magnitudes, so that the $F$ catalog may sample a significantly different galaxy population.

Red bandpass correlation functions for faint galaxies have been published by Stevenson *et al.* (1985), Jones *et al.* (1987), and, at very faint magnitudes, by Efstathiou *et al.* (1991) and Brainerd *et al.* (1994). These studies were limited to small ($\theta \lesssim 0°.5$) angular separations. The red bandpass, $r_F$, employed by both Stevenson *et al.* and Jones *et al.* was a combination of a IIIaF emulsion and a GG630 filter; our $F$ bandpass, IIIaF + GG495, is close enough to $r_F$ to allow a direct comparison of results.

Our $F$ catalog correlation functions are presented in Figure 6 for various limiting magnitudes; fitting parameters for these correlation functions are given in Tables 3 and 4. There is good agreement between the correlation functions for galaxy samples chosen by limiting magnitude or by magnitude range (as found for the $J$ catalog). We have therefore chosen *not* to plot the results for magnitude slices separately; however, fitting parameters for both samples may be found in the table.

Figure 7 superimposes estimates of $w(\theta)$ measured separately in the four fields for which we have $F$ plates; all objects with $F < 23$ are included in this plot. The shape and amplitude of $w(\theta)$ are very similar in all four fields, a conclusion that was also drawn for the $J$ catalog data. The mean field-to-field rms scatter among these 4 fields is $\pm 0.10$ in log $w(\theta)$, or $\pm 0.05$ for the mean log $w(\theta)$ at each separation ($\pm 12\%$ in $w(\theta)$).

It is evident in Figure 6 that the slope $\delta$ becomes shallower towards fainter magnitude limits; this is also apparent in Table 3. The effect is formally significant at the $> 99\%$ level. Furthermore, a similar result is found from fits to the data in individual fields (Fig. 7). As was the case for the $J$ data, we conclude that the flattening of $\delta$ at faint magnitudes is real, at least in a formal statistical sense (systematic errors are discussed further in §5). (Note that the fitted $\delta = 0.94$ for $F < 21$; this value seems unreasonably large when compared to (i) $\delta = 0.79$ for $20 \leq F \leq 21$, and (ii) $\delta = 0.82$ for the corresponding bright $J$ sample ($J < 22$). However, the observed difference is only a $2\sigma$ effect. Furthermore, a fit at small $\theta$ ($\lesssim 0°\!.04$) yields a slope of $0.80 \pm 0.04$, which seems more reasonable. Adopting $\delta = 0.8$ at $F < 21$ still results in a $> 98\%$ significance variation in $\delta$ with magnitude.)

We find that the amplitude $A_w^{\delta=0.8}$ decreases towards fainter magnitudes, as is expected. The mean amplitudes computed for the 4 fields individually are $A_w^{\delta=0.8}(1') = 0.045 \pm 0.008$ ($F \leq 23$), $0.058 \pm 0.013$ ($F \leq 22$), and $0.122 \pm 0.011$ ($F \leq 21$). These determinations are close to the global area results at all limiting magnitudes.

### 4.3 Correlations in the Merged Catalog

We divided the galaxies in our $M$ catalog (which consists of all objects common to both $J$ and $F$ catalogs) into two groups, with $(J - F) < 1.0$ and $(J - F) > 1.0$. The angular correlation function was then computed as a function of limiting magnitude for each color class independently. The results are displayed in Figure 8, and are tabulated in Tables 5 and 6. As in the previous $J$ and $F$ results, there is a tendency for the slope of $w(\theta)$ to decrease towards the fainter magnitudes. The cutoffs and breaks in the color-separated subsamples are identical to within the errors. In addition, the amplitude scales with the number density of galaxies according to the scaling relation (§5). Perhaps the most interesting result is that the amplitude $A_w^{\delta=0.8}$ is almost identical for the blue and red subsamples. This is discussed further in §5 below.

## 5. Discussion

### 5.1 Power Law Index

Most of our power-law index values (Tables 1 and 3) lie roughly between the canonical value of 0.8, and the value of 0.66 found by Maddox et al. (1990b). However, the $J$ and the $F$ data discussed in §4 clearly show that the power law index of $w(\theta)$ decreases as the limiting magnitude increases; such an effect is also seen in the data binned by magnitude slices. The formal significance of the effect (estimated by using a $\chi^2$ statistic to test for constant $\delta$) ranges from $> 75\%$ to $> 99\%$, depending on the subset of data used.

This effect has been seen in some previous work (e.g. Shanks et al. 1980, Koo and Szalay 1984, Pritchet and Infante 1986). Neuschaefer et al. (1991) have also found that $w(\theta)$ becomes shallower at fainter limiting magnitudes, but concluded that the effect was probably not significant in their data. Jones et al. (1987) failed to find such an effect; they concluded that, at all separations and depths, a power law slope of $\delta = 0.8$ was consistent with their data. (The Jones et al. data does not strongly contradict the existence of some variation in $\delta$, because their $w(\theta)$ measurements were made over a fairly limited angular range $[0°\!.002 < \theta < 0°\!.02]$.) Stevenson et al. (1985) state that a slope of $-0.8$ "is at least consistent" with their data at the faintest magnitudes ($J \leq 23$, $F \leq 22$), but using their data it is not possible to rule out slopes as shallow as $-0.65$ in $J$ and $-0.55$ in $F$.

There are several effects in the data or data reduction techniques that might in principle cause changes in the slope of $w(\theta)$ at faint magnitudes.

(i) An incorrect value of $B$, the integral constraint factor, could affect $\delta$. However, our $B$ values are very close to 1; lowering them slightly would have the effect of decreasing $\delta$ still further, as is clear from equation 5 (see also Pritchet and Infante 1986). (Raising $B$ to 1 would have a negligible effect because $B$ is already so close to unity; $B > 1$ is unrealistic in a Universe with a large [and growing] amount of clustering, *unless* there exists a significant amount of negative correlation power on very large scales.) Furthermore, our $B$ values were estimated in a self-consistent manner (see §3), and should be correct unless there exist positive correlations beyond the scale of our plates (in which case $B$ will be overestimated and the true slope will be even flatter).

(ii) Magnitude zeropoint errors across the plate could in principle lead to spurious correlations at large scales. Random variations in zeropoints would produce noise in $w(\theta)$, but not a systematic overestimate at large scale. A more likely possibility is the existence of radial gradients in zeropoints across individual plates. This was investigated in Paper I; based on the mean distribution of galaxy density across the plates, we concluded that there was no evidence for such an effect in our $J$ data, and that such an effect, though marginally present, was probably not significant in our $F$ data. Nevertheless, to test this hypothesis further, we have simulated a radial gradient in magnitude zeropoint in our $J$ data. We find that radial gradients in zeropoint as large as 0.3 mag from plate center to edge have hardly any effect on the slope of the correlation function.

Finally, we note that any population of uncorrelated images (*e.g.* stars, spurious faint images, *etc.*) would lower the correlation amplitudes of galaxies by a factor that would be independent of $\theta$, and hence would not affect $\delta$. We also note that the simulations in Paper II recovered the slope of the correlation function correctly in the mean, even in the presence of irregular field boundaries.

One possible *physical* explanation of the observed effect is that it is due to a real steepening of the slope of the 3D spatial covariance function with increasing cosmic epoch. For example, the simulations of Davis *et al.* (1985) indicate that $d\delta/dz \approx -0.5$ at the present epoch for a cold dark matter initial fluctuation spectrum with $\Omega = 1$, and about $-0.25$ for $\Omega = 0.2$. (Simulations of spatially flat Universes with non-zero cosmological constant give results that are similar to open models with the same value of $\Omega$ – *e.g.* Davis *et al.* [1985], Efstathiou, Sutherland, and Maddox [1990].) If we take a median redshift of $\sim 0.4$ for our $J < 24$ sample (Lilly, Cowie, and Gardner 1991), and $\sim 0.2$ for $J < 22$ (Colless *et al.* 1990, Broadhurst *et al.* 1988), then we would expect a change in $\delta$ of between about $-0.05$ ($\Omega = 0.2$) and $-0.10$ ($\Omega = 1$). This is very roughly consistent with what is observed.

Nevertheless, this consistency should be viewed with some caution. The Davis *et al.* result is for the CDM density field; there is some theoretical evidence that the redshift dependence of the slope (and amplitude) of $w(\theta)$ for *galaxies* is weaker (Carlberg 1991), *if* galaxies form preferentially at high density peaks in the density field. In addition, the predicted shape and evolution of $\xi(r)$ is quite complex. The fitted slope of $\xi(r)$ depends on the range of separations used for the fit, and this behaviour will probably propagate through to $w(\theta)$; galaxies observed at different magnitudes may therefore exhibit slightly different values of the power-law index $\delta$, depending on the median redshift of the sample and the range of angles over which $\delta$ is measured. For Universes with $\Omega < 1$, $\xi(r)$ (and hence $w(\theta)$) may steepen more rapidly at small separations than at large separations (Davis *et al.* 1985). $\xi(r)$ may even *flatten* with time at very small separations, due to mergers (Carlberg 1991, 1993).

An alternate, and perhaps more plausible, explanation for the change in slope with magnitude is due to a dependence of slope with morphological type (Davis and Geller 1976; Giovanelli, Haynes, and Chincarini 1986). This is discussed further below.

## 5.2 Observations of Correlation Amplitudes

Amplitudes were obtained by fitting a power law $w(\theta)$ with fixed slopes of 0.8, 0.66, and 0.6 (Tables 2, 4, and 6). In all cases the points used in the fits corresponded to small separations ($0.001 \lesssim \theta$ [deg] $\lesssim 0.1$), which provides a reasonable match to the data, and is where most other groups have made their fits. We emphasize this point for two reasons. *(i)* The data at small separations possess a reasonably strong amplitude ($w \approx 0.1$ at $\theta = 0°01$), and is hence less affected by any biases that might affect the low amplitude, large separation data. *(ii)* Since different groups fit different slopes to $w(\theta)$, it is important to fit the amplitude in roughly the same angular range to avoid large systematic errors.

We have plotted the $A_w^{\delta=0.8}(1')$ results against $J$ and $F$ magnitude in Fig. 9, and against galaxy surface density in Fig. 10. Figures 9 and 10 also compare our work with results obtained by other groups. The UKST and the AAT data points are both from Stevenson *et al.* (1985), who studied $w(\theta)$ on 1.2 m UKST plates to a limiting magnitude of $b_j < 21$ and $r_F < 20$ for an area $\sim 100$ deg$^2$ near the South Galactic Pole, and on 4m AAT plates to a limiting magnitude $b_j < 24$ and $r_F < 22$ for an area of $\sim 0.2$ deg$^2$. The Koo and Szalay (1984) (hereafter KS) results are for a total area of $\sim 0.2$ deg$^2$. The results of Pritchet and Infante (1986) were obtained from a CFHT plate covering 0.55 deg$^2$ at the South Galactic Pole. Finally, we have also included the recent CCD data of Neuschaefer *et al.* (1991), Efstathiou *et al.* (1991), Roche *et al.* (1993), Couch *et al.* (1993), and Bernstein *et al.* (1994) measured at faint magnitudes.

It can be seen that there is considerable scatter in the various determinations of $A_w$. The scatter is worse in the $J$ band than in $F$. However, our data lies in the middle of various determinations of $w(\theta)$, and joins smoothly onto the amplitudes at brighter and fainter magnitudes. (The continuity with the bright galaxy data from the APM survey, and also from Stevenson *et al.* [1985], is particularly important, because the correlation amplitudes are larger at brighter magnitudes, and hence the correlation function should be determined more accurately.) Furthermore, our work is in good agreement with all *recent* work (except possibly Neuschaefer *et al.* (1991) at the bright end).

### 5.3 Theoretical Interpretation of Correlation Amplitudes

We now turn to a brief discussion of the theoretical interpretation of the amplitude of $w(\theta)$ at small ($\theta \lesssim 0°.1$) separations. (Some of these results have been discussed in preliminary form in Pritchet and Infante [1992].)

The amplitude–magnitude and amplitude–surface density diagrams contain information on the evolution of clustering amplitude with redshift, and also on luminosity evolution and $k$-corrections of galaxies at intermediate redshifts. The effect of cosmology can also be important, especially at the faintest magnitude levels (e.g. Efstathiou et al. 1991). These effects are strongly coupled together, but the effects of luminosity evolution are clearly smaller in the $F$ band than in the $J$ band.

Previous attempts to measure the redshift evolution of clustering amplitude have been inconclusive, not only because of the strong coupling among luminosity evolution, clustering evolution and cosmology mentioned above, but also because of the large scatter in of $A_w$ determined by different observers. Phillipps et al. (1978) and Jones et al. (1987) used the angular correlation function to show that clustering amplitude in proper coordinates was weaker in the past. On the other hand, Loh (1988) used a sample of 1000 galaxies and redshifts determined from narrow-band photometry in the range $0.15 < z < 0.85$; he concluded that clustering is stable in proper coordinates.

From a purely empirical viewpoint, the effects of luminosity evolution on $w(\theta)$ must be very small at $J < 24$. Luminosity evolution enters into the theoretical computation of $w(\theta)$ *only* through the redshift distribution of galaxies, $dN/dz$ (e.g. Phillipps et al. 1978, Peebles 1980, Efstathiou et al. 1991). From extensive work over the past few years it has become apparent that the *observed* redshift distribution of galaxies at $J < 24$ does not differ significantly from that predicted by models without any luminosity evolution (e.g. Broadhurst et al. 1988, Colless et al. 1990, Lilly et al. 1991). It therefore follows that *the amplitude $A_w$ of the angular correlation function for $J < 24$ should be correctly predicted by models without any luminosity evolution.* The same must of course be true in the $F$ band.

It is important to note that simple no-evolution models do *not* correctly predict the number count – magnitude relation (e.g. Maddox et al. 1990a and references therein); hence it is not possible to compare observations and models in the $A_w$ – surface density diagram (Fig. 10). However a comparison of theory and observation in the $A_w$–*magnitude* plane should be unaffected, provided that the observed redshift distribution $dN/dz$ for the observed objects does not differ significantly from the no-evolution model prediction. Also note that the models of Jones et al. (1987) are not appropriate for comparison with our observations, because they have been "tuned" using luminosity evolution to match the observed number counts; this in turn implies that their predicted mean redshift will be too high, and predicted correlation amplitudes $A_w$ too low.

Let us assume a simple model (e.g. Koo and Szalay 1984, Efstathiou et al. 1991) in which the redshift evolution of clustering is represented by a parameter, $\varepsilon$, such that the spatial covariance function is given by

$$\xi(r) = (\frac{r}{r_o})^{-\gamma}(1+z)^{-(3+\varepsilon)} \quad (11)$$

where $r$ is proper distance and $r_o$ is the correlation length. (Davis and Peebles [1983] find $r_0 = 5.4 h^{-1}$ Mpc, and for the moment we assume $\gamma = 1.8$; see §1.2 for justification of these values.) In this equation, $\varepsilon = 0$ corresponds to stable clustering (i.e. "bound" clusters fixed in proper coordinates), $\varepsilon > 0$ to clustering amplitude growing with time in proper coordinates, and $\varepsilon < 0$ to a clustering amplitude decaying with time (in proper coordinates). The specific case $\varepsilon = \gamma - 3 \approx -1.2$ corresponds to clustering that is fixed in comoving coordinates (e.g. as found for galaxies in the biased CDM simulations of Carlberg 1991). (Note that Phillipps et al. [1978] and Jones et al. [1987] instead use a quantity $\beta$ to parametrize the redshift dependence of the spatial correlation function. The quantity $\beta$ is related to $\varepsilon$ by $\beta = \varepsilon/(\gamma - 3) \approx -0.8\varepsilon$.)

As discussed above, the calculation of $w(\theta)$ from $\xi$ depends on the distribution of overlapping structures along the line of sight (i.e. on $dN/dz$). To see this, we note that the correlation function in eq. 11 results in an angular correlation function

$$w(\theta) = A_w \theta^{(1-\gamma)}, \quad (12)$$

where $\theta$ is in radians, and $A_w$ is given by

$$A_w = C r_0^\gamma \frac{\int_0^\infty g(z)(dN/dz)^2 dz}{[\int_0^\infty (dN/dz)dz]^2} \quad (13)$$

(*e.g.* Efstathiou *et al.* 1991; see also Peebles 1980, Phillipps *et al.* 1978). Here $C$ is a constant involving purely numerical factors, *viz.*

$$C = \sqrt{\pi}\frac{\Gamma[(\gamma-1)/2]}{\Gamma(\gamma/2)}. \quad (14)$$

The function $g(z)$, which depends only on $\epsilon$, $\gamma$, and cosmology, is given by

$$g(z) = (\frac{dz}{dx})x^{1-\gamma}F(x)(1+z)^{-(3+\epsilon-\gamma)}, \quad (15)$$

where $x(z)$ is coordinate distance, and $F(x)$ is

$$F(x) = [1 - (H_o a_o x/c)^2(\Omega_o - 1)]^{\frac{1}{2}} \quad (16)$$

(Peebles 1980, eqns. 56.7 and 56.13). The strong dependence of $A_w$ on $dN/dz$ is clear in equation 13. Note that *there is no dependence on galaxy evolution in equation 13, except in the calculation of the redshift distribution $dN/dz$.*

As noted above, it is appropriate to use a no-evolution model for $dN/dz$ to calculate $A_w$; the fact that such a model fails to account for the number counts is irrelevant if attention is restricted to the $A_w$-magnitude plane. Fig. 11 and Table 7 present $A_w$ values computed according to the prescription above for both the $J$ and $F$ bandpasses. The "standard" model (solid line in Fig. 11) is computed using $\gamma = 1.8$, $\Omega = 0.2$, $\varepsilon = 0$, and $r_o = 5.4h^{-1}$ Mpc. (Correlation amplitudes scale as $r_o^\gamma$.) The no-evolution $dN/dz$ is taken from Metcalfe *et al.* (1991), but similar results are obtained using the no evolution models of Broadhurst *et al.* (1988). Our final $A_w$ values are in reasonable agreement with the no-evolution values tabulated by Koo and Szalay (1984) for the $J$ band.

Also shown in Fig. 11 is the $A_w$ that would result from the use of the actual *observed* $dN/dz$ from Colless *et al.* (1990) and Broadhurst *et al.* (1988). This is in very good agreement with calculations from our no-evolution models.

The sensitivity of the theoretical $A_w$ values to cosmology is very small, at least at these magnitude levels. There is no sensitivity to $H_o$, because the quantity $H_o a_o x/c$ is dimensionless, and because the $H_o$ dependence of $g(z)$ is exactly cancelled by that in the measured value of $r_0$. The effect of raising $\Omega$ to 1 is to increase the predicted log $A_w$ by only 0.02–0.06 (depending on passband and magnitude).

There is of course some sensitivity to clustering evolution in our models. Adopting $\varepsilon = -1.2$ raises log $A_w$ by $\sim 0.12$, whereas raising $\varepsilon$ to $+2$ lowers log $A_w$ by $\sim 0.2$ relative to the standard model. (It should be noted that $\varepsilon > 0$ corresponds to a bound clustering pattern that is collapsing in proper coordinates, a point that is discussed further below.)

Finally we consider the dependence of $A_w$ on $\gamma$. Fig. 11 and Table 7 show that that $A_w$ decreases by a factor of 1.6× as $\gamma$ goes from 1.8 to 1.6. (In fact, the *true* decrease in $A_w$ may be slightly less than this. The determination of $\xi(r)$ from redshift surveys such as that of Davis and Peebles (1983) is mostly weighted to points with $r < r_o$, with a typical "mean" $r \simeq$ 2–3 $h^{-1}$ Mpc. Adopting $\gamma = 1.6$ instead of 1.8 would require a larger $r_o$ (by about 17%) to fit the present epoch $\xi(r)$ data, and hence would result in an increase of about 0.07 in log $A_w$. This effect has not been included in Fig. 11, which assumes constant $r_o$ and $\gamma$.)

How does the above model fare at magnitudes brighter than $J \simeq 20$, where the effects of cosmology and clustering evolution are even smaller? Fig. 12 shows a comparison of the APM Survey $w(\theta)$ for $17 < b_J < 20$ (Fig. 1 of Maddox *et al.* 1990*b*) with a model with $r_o = 5.4h^{-1}$ Mpc, and $\gamma = 1 + \delta = 1.66$ and 1.8. (A sharp cutoff in $\xi(r)$ at $r = 25$ Mpc has been used to obtain a cutoff in $w(\theta)$ roughly as observed by Maddox *et al.* ; the nature of this cutoff does not affect our discussion.) The model and observations are in excellent agreement for $\gamma = 1.66$. Using the somewhat steeper slope $\gamma = 1.8$ does not match the data as well, but the overall agreement is still acceptable.

### 5.4 Comparison of $J$ Observations with Models

The "standard model" (no clustering *or* luminosity evolution) is compared with our amplitude observations in Fig. 13. It can be seen that our observations *at all $J$ magnitudes* lie *far* below the predictions of the no-evolution model with $r_0 = 5.4h^{-1}$ Mpc, by about a factor of 4 at $J = 24$ and a factor of 2.5 at $J = 22$. Formally, the observations require $\varepsilon \approx 5$, or $r_0 \approx 3h^{-1}$ Mpc, as shown in the Figure. Lowering $r_0$ to $4.7h^{-1}$ Mpc (Groth and Peebles 1977) would clearly not solve the problem.

Suppose that a shallower slope $\delta = 0.66$ (as found in the APM Survey by Maddox *et al.* 1990*b*) is used in fitting our correlation amplitudes: what effect does this have? A comparison of models and observations is shown in Fig. 14. Clearly a discrepancy is still present between models and observations, by a factor of $\sim 2.5\times$ at $J \gtrsim 23$. Again, this is in sharp contrast with the situation observed for the APM survey data (Fig. 12), for which a standard model with $\delta = 0.66$ provides a very good fit.

### 5.5 Discussion of Low Correlation Amplitudes in the $J$ Band

Several possible explanations of the low clustering amplitudes in the $J$ band have been proposed; these are discussed below.

*(i) Effects of Cosmology.* Efstathiou et al. (1991) discussed the possibility that low clustering amplitudes at $J \approx 26$ could be explained by changes in the assumed cosmological model. However, the low observed clustering amplitudes in Fig. 13 cannot be explained by cosmological effects, because of the relatively low mean redshift of our survey. Adopting $\Omega = 1$ (rather than 0.2) *increases* the predicted value of $w(\theta)$ and so exacerbates the problem. There is no dependence of the predicted $w(\theta)$ on the Hubble constant $H_o$, as discussed above.

*(ii) Contamination by Foreground and Background Objects.* A two component system possesses correlation amplitude $w = f_1^2 w_{11} + f_2^2 w_{22}$, where $f_1$ and $f_2$ are the fractional contributions of components 1 and 2. It therefore follows that $\gtrsim 50\%$ of all galaxies at $J = 24$, or $\gtrsim 35\%$ of objects at $J = 22$, would have to belong to a weakly clustered component (a component not included in the no evolution models) in order to produce the observed correlation amplitudes. Aside from the possibility that the bulk of the galaxies are intrinsically weakly clustered (see item *(iv)* below), there appear to be two possibilities for such a contaminating component: stars, or high redshift galaxies.

The possibility of high redshift galaxies contaminating the survey can be easily disposed of. Approximately 5% of the galaxies ($J < 22.5$) in the Colless et al. (1990) redshift survey remain without redshifts (Colless et al. 1993). It is clear that, even if *all* of these objects are placed in an unclustered population at high redshift, the effect on the empirical dN/dz distribution, and hence on our computed $w(\theta)$ values, will be very small. (The Lilly et al. [1991] $B_{AB} < 24.1$ redshift sample also gives a comparable result, but the statistical uncertainties are larger because only 2 of 22 objects remained unclassified.)

What about stars? At faint magnitudes ($J = 24$) the dilution of $w(\theta)$ would be small because the fraction of stellar objects is around 10% (resulting in at most a 20% underestimate of $w(\theta)$). At $J = 22$, the fraction of stellar objects rises to $\sim 50\%$, but most of these are removed by the star/galaxy classification algorithm. (The fact that this is working as expected at $J = 22$ is confirmed by the excellent agreement between observations and star count models at this magnitude [Infante 1994b].)

*(iii) Models with $\varepsilon > 0$.* In Efstathiou et al. (1991) and earlier papers it was assumed that $-1.2 \leq \varepsilon \leq 0$, with the upper bound corresponding to a clustering pattern that is stable in proper coordinates. However, some recent work (Melott 1992; Yoshii, Peterson, and Takahara 1993) has suggested that $\varepsilon$ can be greater than zero in $n$-body simulations, because new clusters of objects are constantly forming as the Universe expands.

However, there are several reasons why such models cannot explain the observations. First, the models of the above authors all give $\varepsilon < 3$ for the redshift range of interest – significantly less than required ($\varepsilon \approx 5$) to explain the observations. Second, Roukema and Yoshii (1993) have shown that allowing newly formed groups of objects to merge reduces the rate of increase of the correlation function with cosmic epoch – *i.e.* reduces $\varepsilon$. (Nevertheless, the poor match of their models with observations of $w(\theta)$ may be due to their use of redshift distributions including evolution.) Third, and perhaps most important, the use of $\varepsilon = 5$ produces such a rapid evolution of $\xi(r)$ with redshift that there is a significant mismatch between APM observations and models (*e.g.* Fig. 12), by about a factor of $1.6\times$ (even though the mean redshift of the APM sample is only about 0.1).

*(iv) Models with Reduced Correlation Length.* Here we consider the possibility that the bulk of the galaxy population at magnitudes 22–24 in $J$ is intrinsically weakly clustered – *i.e.* possesses a smaller correlation length than the $r_o = 5.4h^{-1}$ Mpc derived locally. (While some local measurements of $r_0$ are a bit smaller than this canonical value of $r_0$ [*e.g.* references in §1.2], none come close to our measured values at intermediate redshift.) This hypothesis is the one preferred by Efstathiou et al. (1991); here it is arrived at in a more model independent way, because the choice of cosmology does not affect the interpretation of low correlation amplitudes, and because the redshift distribution is measured rather than assumed.

Brainerd *et al.* (1994) derive a correlation length $\sim 2h^{-1}$ Mpc for very faint galaxies, with redshifts inferred to be $\sim 1$. While this is well beyond the redshift range covered in our survey, continuity considerations might then be used to argue that galaxies at intermediate redshifts, such as those we have observed, should also have a "lower than local" value of $r_0$ (although not as extreme as $r_0$ at $z = 1$).

(Cole *et al.* [1994] have recently measured $r_0$ from a pencil beam redshift survey $[17 < B < 22]$ to be $6.5h^{-1}$ Mpc. We cannot explain this result, but note that it would result in angular correlation amplitudes significantly in excess of our observations [by $\sim 4\times$], and also of observations by other groups shown in Figs. 9–10. Furthermore, a recent determination of $r_0$ from a faint spectroscopic survey [Shepherd and Carlberg 1994] does not confirm the Cole *et al.* result, but rather finds a value more like 3 $h^{-1}$ Mpc.)

At this point the reader is referred to the discussion in the Introduction relating to the nature of faint ($J \gtrsim 21$), intermediate redshift ($\langle z \rangle \simeq 0.3$) galaxies. There it was concluded that a plausible (though not unique) model for the excess counts and luminosity density at these redshifts involved low-mass galaxies brightened by bursts of star formation (possibly induced by mergers and interactions). If these "dwarfish" objects possess weak clustering properties compared to normal galaxies, then the observations of $w(\theta)$ can be explained naturally.

Over the interval $20 < J < 23$ the observed number of objects is about a factor of $2\times$ larger than would be expected from a no-evolution model (*e.g.* Maddox *et al.* 1990a). For $\delta = 0.66$, the correlation function is observed to be a factor of $\sim 2.5\times$ smaller than expected at this magnitude. It follows that the correlation function of the "excess" population is $\sim \frac{1}{2}$ of the correlation amplitude of "normal" galaxies. This relatively modest difference in correlation power is sufficient to explain the observations, as will be discussed further below. (The correlation power of the "excess" population that we have derived would have to be lowered if there were significant cross-correlation power between the normal and "excess" populations. Such a cross-correlation effect may have been observed locally by Vader and Sandage [1991], and may be explained naturally in the scenario proposed by Babul and Rees [1992].)

It is worthwhile emphasizing that such a model is not as *ad hoc* as it might at first appear. It is known empirically that there are several classes of extragalactic objects that are more weakly clustered than the canonical $r_o = 5.4h^{-1}$ Mpc would predict. IRAS galaxies possess $r_o = 3.8h^{-1}$ Mpc, with a flatter power law slope of $\gamma = 1.57$ (Saunders, Rowan-Robinson, and Lawrence 1992; Strauss *et al.* 1992; Moore *et al.* 1994). Gas-rich low mass galaxies with young stellar populations ("H II Galaxies") possess $r_o = 2.7h^{-1}$ Mpc (Iovino, Melnick, and Shaver 1988). Ly$\alpha$ absorbers appear to be more weakly correlated with galaxies than galaxies are with themselves (Morris *et al.* 1993). Davis and Geller (1976) and Giovanelli *et al.* (1986) both find that later morphological types are significantly more weakly clustered; for example Giovanelli *et al.* find that Sc and later galaxies have correlation amplitudes $\sim 4 - 5\times$ weaker than E/S0 galaxies. A similar effect is also found by Babul and Postman (1990). Davis and Djorgovski (1985) find that low surface brightness galaxies are more weakly correlated than high surface brightness galaxies.

Furthermore, two of these classes of objects (IRAS sources and emission-line galaxies) may be generically related to the "excess" population at intermediate redshift, since starbursts have also been used to explain these objects. Thus we have an empirical link between the correlation properties of galaxies at intermediate redshift and the correlation function of supposedly similar objects at low redshift. (A simpler explanation could simply be that the faint galaxy population possesses the correlation properties found locally for galaxies later than S0/a (Babul and Postman 1990). From this point of view there is nothing whatsoever abnormal about the clustering properties of faint galaxies [out to $z \approx 0.5$]!)

Remarkably, *all* of the objects listed above that possess weak correlation amplitudes *also* possess lower values of $\delta$ – *i.e.* flatter power-law correlation functions. For example, Giovanelli *et al.* (1986) find that $\delta \simeq 0.4$ for late-type spirals (Sbc and later and irregulars, compared to $\delta \simeq 0.65 - 0.8$ for early-type spirals, and $\delta \simeq 0.8 - 1.0$ for E/S0's. A similar effect is seen in the work of Babul and Postman (1990). This effect is large enough to completely explain the flattening of our correlation functions at faint magnitudes.

Biased cold dark matter models for the development of structure in the Universe predict that the clustering amplitude of dwarf galaxies should be weaker than that of more massive objects (*e.g.* Dekel and Silk 1986), because dwarfs are hypothesized to originate from lower amplitude peaks in the primordial fluctuation spectrum. An alternate scenario that produces less clustered dwarf galaxies is that of Babul and Rees (1992), in which the first burst of star formation in *field* dwarf ellipticals is quenched by a supernova-driven wind, resulting in these galaxies fading to obscurity (perhaps consistent with the model of Cowie *et al.* 1991). (This is in contrast to dE galaxies in clusters, which retain their gas and undergo repeated star bursts, because of a dense intracluster medium.)

The observational detection of such an effect is, however, fraught with difficulties, as has been discussed by many authors. Salzer, Hanson, and Gavazzi (1990) use three separate statistics ($\xi(r)$, nearest neighbor tests, and space density calculations) to demonstrate a difference in the clustering strength of "low luminosity" ($M_B > 18$) and high luminosity ($M_B < -18$) galaxies. They find that high luminosity galaxies are roughly twice as strongly clustered as low luminosity galaxies; this result is in good agreement with the predictions of biased CDM models (White *et al.* 1987), and also in qualitative agreement with our low correlation amplitudes above, if these are due to contamination of faint galaxy surveys with "luminosity-boosted" low mass galaxies (Broadhurst *et al.* 1988, Lilly *et al.* 1991, Cowie *et al.* 1991).

A number of independent studies have corroborated the existence of lower correlation amplitudes for dwarf galaxies (*e.g.* Davis and Djorgovski 1985; Giovanelli et al. 1986; Giovanelli and Haynes 1988; Hollósi and Efstathiou 1988). However, other studies have failed to find such an effect (*e.g.* Thuan, Gott and Schneider 1987, Eder *et al.* 1989). At present it is difficult to say with certainty that this effect exists in low redshift samples; further observations are clearly required.

Since, in biased CDM models, the low mass galaxies provide a more accurate tracer of the true mass distribution in the Universe, it follows that the slope of $\xi(r)$ for dwarf galaxies should be closer to $\xi(r)$ for the dark matter distribution, and flatter than that for $\sim L^*$ galaxies. The difference in the slope of $\xi(r)$ (or, equivalently, $w(\theta)$) between massive galaxies and the CDM substrate is highly model dependent, but may be $\sim 0.2 - 0.4$ (Carlberg 1991, Davis *et al.* 1985). Thus we have an alternative explanation of the flattening of the slope of $w(\theta)$ for faint galaxies: a significant fraction of such objects are low luminosity galaxies that are not found locally, with weaker clustering amplitudes than massive galaxies, and with a flatter correlation slope, as predicted by biased CDM models.

## 5.6 Comparison of $F$ Observations with Models

We now turn to a discussion of the correlation amplitudes in the $F$ band (Fig. 15). The effects of luminosity evolution must be lower in $F$ than in $J$, so the no-evolution redshift distribution dN/dz should be even more applicable to the $F$ data. (In fact this has been confirmed by Lilly (1993): the redshift distribution of galaxies selected in the red matches no-evolution predictions very well.) No-evolution models (calculated as described above for the $J$ band) are plotted in Fig. 11*b*. The effects of cosmology, clustering evolution, and $\gamma$ are very similar to those for the $J$ band, and so will not be discussed separately.

It is clear that, as found for the $J$ band, the no-evolution model does a poor job of matching the observed correlation amplitudes. The no-evolution model predicts correlation amplitudes a factor of $\sim$ 1.6 higher than observations at $F = 21$, rising to a factor of $\sim 2$ at $F = 23$. This discrepancy is not as large as was the case for the $J$ band, but it is very significant. Using $\delta = 0.66$ rather 0.8 (Fig. 16) does not remove the discrepancy. At least some redshift evolution in clustering amplitude would appear to be necessary, as was found for the $J$ data. Alternately, and more reasonably, it is necessary to hypothesize the existence of a population of objects that possesses low clustering amplitude. It may be possible to identify this population with the excess objects in the faint source counts (Lilly *et al.* 1991), as was the case for the $J$ data.

The fact that the $F$ band correlation amplitudes are closer to the no-evolution model than the $J$ band amplitudes indicates that a smaller fraction of galaxies observed in $F$ belong to a weakly clustered population. In other words, our observations in both $J$ and $F$ are consistent with a model in which the bulk of the objects at $J < 24$ consist of a weakly clustered population with bluer than average colors. This is discussed further below.

## 5.7 Color

It is well known that galaxies at faint magnitudes tend to become intrinsically bluer (*cf.* Tyson 1988, Lilly *et al.* 1991; Paper I; Kron 1980; Koo 1986; Infante, Pritchet, and Quintana 1986). The nature of these blue objects remains unclear. On the one hand, models by Bruzual (*e.g.* Figure 8 of Bruzual 1988) suggest that most blue galaxies with $J \lesssim 23.5$ are at $z > 0.5$. On the other hand, this appears incompatible with the $z$ distributions of Colless *et al.* (1990) and Lilly *et al.* (1991).

In §4 we discussed the amplitude of the cross-correlation function for the merged $J$ and $F$ catalog sample as a function of color. The principal result from this section was that the correlation amplitude for blue and red galaxies at a given $(J+F)/2$ was identical to within the errors. However, in the previous section we noted that the correlation amplitudes for galaxies observed in the $J$ band were weaker (with respect to no-evolution models) than for the $F$ band. This latter result seems to predict that, at fixed $(J+F)/2 \approx V$ magnitude, blue galaxies should be more weakly correlated than red galaxies – in contradiction with the observations discussed in §4 (*cf.* Fig. 8).

The resolution of this apparent paradox appears to lie in the redshift distribution of blue and red galaxies. At a fixed apparent magnitude, for non-evolving or weakly evolving galaxies, blue galaxies possess a redshift distribution with a shape skewed to low redshifts relative to red galaxies. (This is due primarily to $k$ corrections, but also to the color–luminosity relation, which predicts that intrinsically faint galaxies, which are relatively nearby in a magnitude-limited sample, are also bluest. This effect is quite apparent in color-redshift diagrams – *e.g.* Fig. 3 of Koo and Kron 1992.) If all galaxies in the Universe possessed uniform clustering properties, and were at most weakly evolving, then the angular correlation function of blue galaxies would naturally be stronger than that of red galaxies, at fixed apparent magnitude. Fig. 17 demonstrates this effect for galaxies in the magnitude range $20 < (J+F)/2 < 22.5$. The no-evolution model predicts a median redshift for galaxies with $J-F < 1$ of 0.23, compared to 0.35 for $J-F > 1$. This has a big effect on correlation amplitudes: the predicted angular correlation amplitudes would be a factor of about $1.6-2\times$ stronger for blue galaxies $((J-F) < 1)$, *even if* blue galaxies and red galaxies possessed the same *spatial* correlation functions. The fact that our observations show roughly equal *angular correlation* amplitudes for these two populations is therefore consistent with the fact that galaxies selected in blue light seem to be intrinsically more weakly clustered than those selected in red light.

## 5.8 A Simple Model

We conclude the discussion of correlation amplitudes with a simple empirical model that can be used to understand our observations.

The observed color distributions of galaxies are shown in Fig. 18a for the magnitude ranges $20 < J < 23$, $20 < (J+F)/2 < 22.5$, and $19 < F < 22$; these data are taken from Infante and Pritchet (1992). We assume that the reddest galaxies belong to a non-evolving population, and hence scale the no evolution model to match the observed numbers of red galaxies $(J-F \gtrsim 1.6)$ in Fig. 18a (see also Colless *et al.* 1990). (The scaling factor in the $J$ band is about 0.55; it is interesting to note that applying this scaling factor to the luminosity function normalization of Metcalfe *et al.* [1991] reproduces the APM number counts [Maddox *et al.* 1990a] at $b_J < 17$ almost exactly. Fainter than this, the well-known excess of galaxies relative to a no evolution model starts to appear [*e.g.* Fig. 2 of Maddox *et al.*].)

It is clear from Fig. 18a that the excess population of galaxies is significantly bluer than a no-evolution model. The difference between the no-evolution model and the observations is shown as a dashed line in this figure. Comparison of the color distribution of this excess with the color distribution of the five morphological components used in the Metcalfe *et al.* (1991) model (Fig. 18b) demonstrates that the excess population is approximately as blue as the bluest model component ("Sdm" in color, although *not* necessarily in morphology).

The model scaling using Fig. 18a can be used to estimate the fraction of galaxies in the excess population; this fraction ranges from 53% for $20 < J < 23$ to 28% for $19 < F < 22$. Based on these color distributions, we conclude that a population of "Sdm"-like galaxies $\sim 5\times$ larger than observed locally would explain the excess counts and the color distributions in the above magnitude ranges. This is confirmed in Fig. 19, which compares the model color distributions (after adding the excess population) with the observations. The agreement is remarkably good.

Finally, we assume that the correlation function of the excess population is about $2.5\times$ lower than that of "normal" galaxies (*i.e.* $r_o = 3.2$ Mpc rather than the canonical 5.4 Mpc, if $\gamma = 1.8$). This results in an observed angular correlation amplitude about $3\times$ lower at $J = 23$, and $2\times$ lower at $F = 22$, roughly as observed (*e.g.* Figs. 13 and 15). Furthermore, this model predicts approximately equal correlation amplitudes (to within 0.1 in log $w(\theta)$) for a samples of objects with $20 < V < 22$ cut according to whether $J - F < 1$ or $> 1$. Thus this model reproduces all of our observations.

This simple model makes a number of assumptions, is not optimized in any way, and is certainly not unique. Principal among the assumptions made is that the excess population possesses a luminosity function identical to that assumed for Sdm galaxies locally by Metcalfe et al. (1991); however, if this were not the case, then the redshift distribution of galaxies would differ markedly from that predicted by a no-evolution model. (The fact that the colors of the excess population are the same as that of Sdm model component does not necessarily mean that the luminosity functions are the same.) The model ignores any cross-correlation power between excess and normal components. The assumption of a single component for the excess galaxies is probably unrealistic. Almost certainly there remain significant uncertainties in the type dependent luminosity functions. Nevertheless, the model demonstrates that it *is* indeed possible to explain, in a roughly self-consistent manner, the low observed correlation amplitudes as being due to a *weakly-clustered, blue* population.

## 6. Conclusions

The principal conclusion of this paper is that faint galaxies are more weakly clustered than galaxies observed at the present epoch. There are several classes of galaxies observed locally (for example, starburst galaxies and H II region galaxies) that also possess weaker than average clustering; given the fact that starbursts have been invoked to explain the numbers and colors of faint galaxies, it is tempting to link the clustering properties (amplitude *and* power-law slope) of faint galaxies to the starburst phenomenon. A simple model, in which the "excess" population of weakly-clustered faint blue galaxies possesses colors like those of Sdm galaxies locally, can explain our observations. However, it should be borne in mind that the clustering properties of starburst galaxies remain unexplained at the present time. A simpler explanation may be that faint galaxies are drawn from a population related to nearby late-type galaxies, for which there exists some evidence of weak clustering.

It would be extremely interesting to extend our survey of the clustering of faint galaxies to the $U$ and $K$ bands to further constrain our simple model; observations of $w(\theta)$ at smaller pairwise separations would help us to understand the role of mergers in galaxy evolution (see Carlberg et al. 1994), and would perhaps allow us to indirectly study the formation of disks (Cowie et al. 1993). Surveys that cover even larger solid angles are required to test the hypothesis that faint galaxies have a cutoff in their correlation function at relatively short scales ($\sim 5h^{-1}$ Mpc).

An enormous amount of work remains to be done on the properties of galaxies at the present epoch. The clustering properties of such objects as dwarf galaxies, starburst galaxies, and H II region galaxies are still very poorly known, as is the morphological type dependence of the correlation function. Our knowledge of luminosity functions (particularly with type-dependence) is in a very rudimentary state. Clearly the properties of local galaxies must be better understood if we are to make progress in understanding galaxies at intermediate redshift.

We are grateful to Ray Carlberg, Richard Ellis, David Hartwick, and Rick White for useful discussions. This work was started as a Ph.D. dissertation at the University of Victoria by LI, who gratefully acknowledges financial support from a University of Victoria Fellowship and Petrie Fellowship. CJP acknowledges financial support from the Natural Sciences and Engineering Research Council of Canada and the University of Victoria, and LI thanks Fondecyt Chile for support through Projecto 1930570.

---

This 2-column preprint was prepared with the AAS LaTeX macros v3.0.

Table 1: Fits to $J$ Correlations

| Type | N | B | $A_w(1')$ | $\delta$ | $\theta_c(°)$ | $\log(\theta_b)$ |
|---|---|---|---|---|---|---|
| $J \leq 24$ | 39643 | 0.999 | 0.030 | 0.66 | 0.5 | -0.9 |
| | | | ±0.009 | ±0.05 | ±0.2 | |
| $J \leq 23$ | 13548 | 0.997 | 0.051 | 0.78 | 1.2 | -0.6 |
| | | | ±0.007 | ±0.03 | ±0.4 | |
| $J \leq 22$ | 4811 | 0.994 | 0.109 | 0.82 | > 2 | -0.4 |
| | | | ±0.017 | ±0.05 | | |
| $23 \leq J \leq 24$ | 26132 | 0.997 | 0.020 | 0.69 | > 2 | -0.5 |
| | | | ±0.003 | ±0.03 | | |
| $22 \leq J \leq 23$ | 8787 | 0.995 | 0.044 | 0.71 | > 2 | -0.7 |
| | | | ±0.011 | ±0.05 | | |
| $21 \leq J \leq 22$ | 3419 | 0.991 | 0.107 | 0.79 | 0.5 | -0.5 |
| | | | ±0.022 | ±0.06 | ±0.1 | |

Table 2: Fixed $\delta$ Fits to $J$ Correlations

| Type | $A_w^{\delta=0.8}(1')$ | $A_w^{\delta=0.66}(1')$ | $A_w^{\delta=0.6}(1')$ |
|---|---|---|---|
| $J \leq 24$ | 0.030±0.002 | 0.032±0.002 | 0.033±0.002 |
| $J \leq 23$ | 0.045±0.004 | 0.049±0.003 | 0.050±0.003 |
| $J \leq 22$ | 0.106±0.008 | 0.110±0.006 | 0.111±0.007 |

Table 3: Fits to $F$ Correlations

| Type | N | B | $A_w(1')$ | $\delta$ | $\theta_c(°)$ | $\log(\theta_b)$ |
|---|---|---|---|---|---|---|
| $F \leq 23$ | 38912 | 0.998 | 0.054 | 0.65 | 0.5 | -0.9 |
| | | | ±0.013 | ±0.05 | ±0.1 | |
| $F \leq 22$ | 16712 | 0.992 | 0.080 | 0.66 | 0.3 | -1.0 |
| | | | ±0.013 | ±0.04 | ±0.4 | |
| $F \leq 21$ | 6796 | 0.993 | 0.117 | 0.94 | 0.6 | -0.5 |
| | | | ±0.019 | ±0.04 | ±0.3 | |
| $22 \leq F \leq 23$ | 22337 | 0.997 | 0.060 | 0.60 | 0.5 | -0.9 |
| | | | ±0.016 | ±0.05 | ±0.1 | |
| $21 \leq F \leq 22$ | 9992 | 0.996 | 0.064 | 0.66 | 1.1 | -0.7 |
| | | | ±0.013 | ±0.04 | ±0.1 | |
| $20 \leq F \leq 21$ | 4365 | 0.994 | 0.081 | 0.79 | 1.5 | -0.5 |
| | | | ±0.018 | ±0.06 | ±0.1 | |

Table 4: Fixed $\delta$ Fits to $F$ Correlations

| Type | $A_w^{\delta=0.8}(1')$ | $A_w^{\delta=0.66}(1')$ | $A_w^{\delta=0.6}(1')$ |
|---|---|---|---|
| $F \leq 23$ | 0.053±0.004 | 0.057±0.003 | 0.059±0.003 |
| $F \leq 22$ | 0.077±0.007 | 0.083±0.004 | 0.085±0.003 |
| $F \leq 21$ | 0.132±0.008 | 0.137±0.012 | 0.139±0.014 |

Table 5: Fits to Merged Catalog Correlations

| $(J+F)/2$ | $J-F$ | N | B | $A_w(1')$ | $\delta$ | $\theta_c(°)$ |
|---|---|---|---|---|---|---|
| 23 | 0-1 | 5822 | 0.987 | 0.083 | 0.54 | 0.7 |
| | | | | ±0.012 | ±0.03 | ±0.2 |
| | 1-2 | 10068 | 0.990 | 0.100 | 0.48 | 0.7 |
| | | | | ±0.028 | ±0.05 | ±0.2 |
| 22 | 0-1 | 2038 | 0.994 | 0.147 | 0.55 | 0.3 |
| | | | | ±0.034 | ±0.05 | ±0.3 |
| | 1-2 | 4373 | 0.994 | 0.135 | 0.73 | 0.6 |
| | | | | ±0.04 | ±0.07 | ±0.01 |
| 21 | 0-1 | 730 | 0.999 | 0.300 | 0.80 | – |
| | | | | ±0.056 | ±0.01 | |
| | 1-2 | 1486 | 0.989 | 0.305 | 0.75 | 0.6 |
| | | | | ±0.073 | ±0.05 | ±0.03 |

Table 6: Fixed $\delta$ Fits to Merged Catalog Correlations

| $(J+F)/2$ | $J-F$ | $A_w^{\delta=0.8}(1')$ | $A_w^{\delta=0.66}(1')$ | $A_w^{\delta=0.6}(1')$ |
|---|---|---|---|---|
| 23 | 0-1 | 0.082±0.012 | 0.080±0.007 | 0.079±0.006 |
| | 1-2 | 0.109±0.016 | 0.106±0.010 | 0.104±0.008 |
| 22 | 0-1 | 0.152±0.024 | 0.143±0.018 | 0.139±0.016 |
| | 1-2 | 0.148±0.012 | 0.139±0.011 | 0.136±0.013 |
| 21 | 0-1 | 0.349±0.065 | 0.279±0.040 | 0.254±0.032 |
| | 1-2 | 0.320±0.021 | 0.311±0.024 | 0.307±0.030 |

Table 7: Predictions of the No-Evolution Model

| mag | $\gamma$ | $\varepsilon$ | $\Omega$ | $A_w(1')$ |
|---|---|---|---|---|
| *(a) "Standard" model* | | | | |
| $17 < J < 20$ | 1.8 | 0.0 | 0.2 | 0.781 |
| $20 < J < 22$ | | | | 0.264 |
| $20 < J < 23$ | | | | 0.175 |
| $20 < J < 24$ | | | | 0.112 |
| $19 < F < 21$ | | | | 0.201 |
| $19 < F < 22$ | | | | 0.143 |
| $19 < F < 23$ | | | | 0.102 |
| *(b) Some variations on "standard" model* | | | | |
| $17 < J < 20$ | 1.66 | 0.0 | 0.2 | 0.494 |
| $17 < J < 20$ | 1.66 | $-1.2$ | 0.2 | 0.563 |
| $20 < J < 22$ | 1.66 | 0.0 | 0.2 | 0.179 |
| $20 < J < 22$ | 1.8 | $-1.2$ | 0.2 | 0.333 |
| $20 < J < 22$ | 1.8 | 0.0 | 1.0 | 0.279 |
| $20 < J < 23$ | 1.66 | 0.0 | 0.2 | 0.121 |
| $20 < J < 23$ | 1.6 | 0.0 | 0.2 | 0.104 |
| $20 < J < 23$ | 1.8 | $-1.2$ | 0.2 | 0.229 |
| $20 < J < 23$ | 1.8 | 2.0 | 0.2 | 0.114 |
| $20 < J < 23$ | 1.8 | 0.0 | 1.0 | 0.187 |
| $20 < J < 23$ | 1.66 | 0.0 | 1.0 | 0.128 |
| $20 < J < 23$ | 1.6 | 0.0 | 1.0 | 0.110 |
| $20 < J < 24$ | 1.66 | 0.0 | 0.2 | 0.079 |
| $20 < J < 24$ | 1.6 | 0.0 | 0.2 | 0.068 |
| $20 < J < 24$ | 1.8 | $-1.2$ | 0.2 | 0.155 |
| $20 < J < 24$ | 1.8 | 0.0 | 1.0 | 0.124 |
| $19 < F < 21$ | 1.66 | 0.0 | 0.2 | 0.140 |
| $19 < F < 21$ | 1.8 | $-1.2$ | 0.2 | 0.266 |
| $19 < F < 21$ | 1.8 | 0.0 | 1.0 | 0.217 |
| $19 < F < 22$ | 1.66 | 0.0 | 0.2 | 0.100 |
| $19 < F < 22$ | 1.6 | 0.0 | 0.2 | 0.087 |
| $19 < F < 22$ | 1.8 | $-1.2$ | 0.2 | 0.197 |
| $19 < F < 22$ | 1.8 | 2.0 | 0.2 | 0.086 |
| $19 < F < 22$ | 1.8 | 0.0 | 1.0 | 0.158 |
| $19 < F < 22$ | 1.66 | 0.0 | 1.0 | 0.110 |
| $19 < F < 22$ | 1.6 | 0.0 | 1.0 | 0.096 |
| $19 < F < 23$ | 1.66 | 0.0 | 0.2 | 0.073 |
| $19 < F < 23$ | 1.6 | 0.0 | 0.2 | 0.064 |
| $19 < F < 23$ | 1.8 | $-1.2$ | 0.2 | 0.148 |
| $19 < F < 23$ | 1.8 | 0.0 | 1.0 | 0.116 |

Note: All models are computed with $r_o = 5.4 h^{-1}$ kpc.

# Figure Captions

FIG. 1 – The distribution of galaxies ($J \leq 23$) for the $2° \times 2°$ NGP catalog (Infante and Pritchet 1992). $\alpha$ and $\delta$ are in degrees from the center of Field 5. Only points inside the contour are included in the analysis.

FIG. 2 – Galaxy pair distribution as a function of angular separation. This distribution is shown for both the $J$ and $F$ catalogs ($20 < J < 24$ and $19 < F < 23$); it is normalized to the total number of pairs in the catalog ($\sim 10^9$).

FIG. 3 – Correlation function of galaxies, $w$, as a function of angular separation $\theta$, for the $J$ catalog. Results for three different limiting magnitudes are shown: (a) $J < 24$; (b) $J < 23$; and (c) $J < 22$. The points represent individual estimates of $w$, and the error bars represent the $\pm 1\sigma$ uncertainty from equation 8. (No attempt has been made to include the other sources of error that were discussed in §3.) The solid line is a fit to the data as described in the text. The dashed line in these and subsequent figures represents a power law with slope $\delta = 0.8$ and amplitude $A_w(1') = 0.080$ (or $A_w(1°) = 0.003$).

FIG. 4 – Correlation function of galaxies, $w$, as a function of angular separation $\theta$, for the $J$ catalog. This figure differs from Figure 3 in that the data is subdivided into magnitude intervals: (a) $23 < J < 24$; (b) $22 < J < 23$; and (c) $21 < J < 22$. See the caption to Figure 3 for further details.

FIG. 5 – Correlation function of galaxies, $w$, as a function of angular separation $\theta$, for the 5 individual fields in the $J \leq 24$ catalog. Here the data for the 5 fields have been superimposed to render field-to-field differences more visible. *Open squares:* field 1; *asterisks:* field 2; *open circles:* field 3; *solid triangles:* field 4; *solid squares:* field 5. The *solid line* is the fit to the data of all 5 fields combined (Fig. 3a). The *dashed line* possesses slope $\delta = 0.8$ and amplitude $A_w(1') = 0.080$.

FIG. 6 – Correlation function of galaxies, $w$, as a function of angular separation $\theta$, for the $F$ catalog. Results for three different limiting magnitudes are shown: (a) $F < 23$; (b) $F < 22$; and (c) $F < 21$. See caption to Figure 3 for more details.

FIG. 7 – Correlation function of galaxies, $w$, as a function of angular separation $\theta$, for the 4 individual fields in the $F \leq 23$ catalog. Here the data for the 4 fields has been superimposed to render field-to-field differences more visible. Symbols have the same meaning as in Fig. 5. The *solid line* is the fit to the data of all 4 fields combined (Fig. 6a). The *dashed line* possesses slope $\delta = 0.8$ and amplitude $A_w(1') = 0.080$.

FIG. 8 – Correlation function of galaxies, $w$, as a function of angular separation $\theta$, for the M catalog, in magnitude and color bins. (a) $(J+F)/2 < 23$; (b) $(J+F)/2 < 22$; and (c) $(J+F)/2 < 21$. In each of these figures, *open circles* refer to the correlation function of red galaxies ($1 \leq (J-F) \leq 2$), and *closed circles* represent the correlation function of blue galaxies ($0 \leq (J-F) \leq 1$). The dashed line possesses slope $\delta = 0.8$ and amplitude $A_w(1') = 0.080$.

FIG. 9 – Correlation function amplitudes ($A_w$) plotted against limiting magnitude. *(a)* $J$ catalog; *(b)* $F$ catalog. All amplitudes are for $\theta = 1'$, and have been corrected to the value that would have been obtained with $\delta = 0.8$. The solid line is the standard model that appears in Fig. 11. See text for further details.

FIG. 10 – Correlation function amplitudes, $A_w$, plotted against surface density of galaxies $N$ (in units of deg$^{-2}$ mag$^{-1}$). *(a)* $J$ catalog; *(b)* $F$ catalog. See text for further details.

FIG. 11 – Predicted correlation amplitudes, $A_w$, against limiting magnitude. (a) $J$ magnitude. (b) $F$ magnitude. The "standard" model (*heavy solid line*) is computed using $\gamma = 1.8$, $\Omega = 0.2$, $\varepsilon = 0$, $r_o = 5.4h^{-1}$ Mpc, with a no-evolution redshift distribution dN/dz from Metcalfe *et al.* (1991). Other lines in the figure show the results of modifying the "standard model". *Dashed line:* $\Omega = 1$; *dotted lines:* $\varepsilon = -1.2$ and 2, with $\varepsilon = -1.2$ lying above the standard model; *thin solid line:* $\gamma = 0.66$. The solid square shows the $A_w$ that results from the observed dN/dz for faint galaxies (obtained by combining the data from Colless *et al.* [1990] and Broadhurst *et al.* [1988]).

FIG. 12 – Angular correlation function measured for $17 < b_J < 20$ in the APM Survey (Maddox et al. 1990b). (The $b_J$ magnitude system is very similar to $J$.) The observations (squares) are compared with models with $r_o = 5.4h^{-1}$ Mpc, and $\gamma = 1.66$ (solid line) and $\gamma = 1.8$ (dashed line). The cutoff in $w(\theta)$ has (somewhat arbitrarily) been obtained by introducing a sharp cutoff in $\xi(r)$ at $r = 25$ Mpc. The dotted line is the correlation function that would result, at this magnitude level, by using an $r_o$ value of $3h^{-1}$ Mpc; this latter value of $r_o$ is close to that of galaxies with $J \gtrsim 22$. The dot-dash line uses $r_o = 5.4h^{-1}$ Mpc and $\varepsilon = +5$.

FIG. 13 – Correlation amplitudes in the $J$ band for $\delta = 0.8$. Points are observations; the solid line is the "standard model" ($r_o = 5.4h^{-1}$ Mpc, $\varepsilon = 0$). The dashed line represents a model with $r_o = 3h^{-1}$ Mpc ($\varepsilon = 0$), whereas the dotted line represents a model with $\varepsilon = +5$ ($r_o = 5.4h^{-1}$ Mpc).

FIG. 14 – Correlation amplitudes in the $J$ band for $\delta = 0.66$. Points are observations; the solid line is the "standard model", and the dashed line and solid line are as in Fig. 13.

FIG. 15 – Correlation amplitudes in the $F$ band for $\delta = 0.8$. Points are observations; the solid line is the "standard model", and the dashed line and solid line are as in Fig. 13.

FIG. 16 – Correlation amplitudes in the $F$ band for $\delta = 0.66$. Points are observations; the solid line is the "standard model", and the dashed line and solid line are as in Fig. 13.

FIG. 17 – No-evolution redshift distributions dN/dz for $20 < (J + F)/2 < 22.5$. Solid line: red galaxies ($(J - F) > 1$); dotted line: blue galaxies ($(J - F) < 1$). The relative vertical normalization of these curves is arbitrary, but does not affect the computation of correlation amplitudes.

FIG. 18 – (a) Observed (thin solid curve) and modelled (thick solid curve) color distributions for $20 < J < 23$, $20 < (J + F)/2 < 22.5$, and $19 < F < 22$. The model is a no-evolution model, scaled to match the numbers of objects for $J - F \gtrsim 1.6$. The dashed curve represents the excess population (difference between observations and scaled no-evolution model). The vertical axis represents the numbers of objects in a 0.1 mag bin in $J - F$, for a solid angle of 1.745 deg$^2$. It can be seen that the excess population is significantly bluer than the no-evolution model. (b) Color distributions for the five components that make up the Metcalfe et al. (1991) model, for the magnitude interval $20 < (J+F)/2 < 22.5$. From right to left, these components correspond to E/S0, Sab, Sbc, Scd, and Sdm. The color distributions have been smoothed to avoid discreteness effects in the color and redshift distributions. Note that the color of the Sdm component is very similar to that of the excess population shown in (a).

FIG. 19 – Model color distribution (thick solid curve) after scaling the Sdm component by 5× relative to the Metcalfe et al. (1991) no-evolution model. The thin solid curve represents the observations in a solid angle of 1.745 deg$^2$, as in Fig. 18. The agreement between model and observations is much better.